\documentclass[11pt,letterpaper]{article}

\usepackage{doublespace}

\usepackage[round,colon]{natbib}
\usepackage{amsfonts,amssymb,amsmath,graphicx,epsfig,amscd,ulem}
\usepackage{graphicx}
\usepackage{array}

\def\bi{\begin{itemize}}

\def\ei{\end{itemize}}
\def\be{\begin{equation}}
\def\ee{\end{equation}}
\def\bea{\begin{eqnarray}}
\def\eea{\end{eqnarray}}

\def\gdot{\dot\gamma}

\DeclareTextSymbol{\degre}{OT1}{23}

\begin{document}

\author{Fabien Mahaut, Xavier Chateau, Philippe Coussot, Guillaume Ovarlez\footnote{corresponding author: guillaume.ovarlez@lcpc.fr}
\footnote{Support from the Agence
Nationale de la Recherche (ANR) is acknowledged (grant
ANR-05-JCJC-0214).}
\\
Universit\'e Paris Est - Institut Navier \\ Laboratoire des
Mat\'eriaux et Structures du G\'enie Civil (UMR 113
LCPC-ENPC-CNRS)\\ 2, all\'ee Kepler, 77420 Champs-sur-Marne,
France }

\title{Yield stress and elastic modulus\\ of suspensions of noncolloidal particles in yield stress fluids}
\maketitle

\renewcommand{\abstractname}{Synopsis}
\begin{abstract}
We study experimentally the behavior of isotropic suspensions of
noncolloidal particles in yield stress fluids. This problem has
been poorly studied in the literature, and only on specific
materials. In this paper, we manage to develop procedures and
materials that allow focusing on the purely mechanical
contribution of the particles to the yield stress fluid behavior,
independently of the physicochemical properties of the materials.
This allows us to relate the macroscopic properties of these
suspensions to the mechanical properties of the yield stress fluid
and the particle volume fraction, and to provide results
applicable to any noncolloidal particle in any yield stress fluid.
We find that the elastic modulus/concentration relationship
follows a Krieger-Dougherty law, and show that the yield
stress/concentration relationship is related to the elastic
modulus/concentration relationship through a very simple law, in
agreement with results from a micromechanical analysis.
\end{abstract}

\section{Introduction}

Dense suspensions arising in industrial processes (concrete
casting, drilling muds, foodstuff transport...) and natural
phenomena (debris-flows, lava flows...) often involve a broad
range of particle sizes. The behavior of these materials reveal
many complex features which are far from being understood (for a
recent review, see \citet{Stickel2005}). This complexity
originates from the great variety of interactions between the
particles (colloidal, hydrodynamic, frictional, collisional...)
and of physical properties of the particles (volume fraction,
deformability, sensitivity to thermal agitation, shape,
buoyancy...) involved in the material behavior.

Basically, these materials exhibit a yield stress and have a solid
viscoelastic behavior below this yield stress; above the yield
stress they behave as liquids, and their flow behavior is usually
well fitted to a Herschel-Bulkley law [\citet{Larson1999}]
although the exact details of the constitutive law seem more
complex at the approach of the transition between the liquid and
the solid regimes [\citet{Coussot2005}]. The yielding behavior
originates from the colloidal interactions which create a jammed
network of interacting particles [\citet{Larson1999,Coussot2005}].
If the behavior of dense colloidal suspensions, and more generally
of yield stress fluids, have received considerable interest and
been widely studied, the influence of the large particles on this
behavior have been poorly studied. Moreover, the few existing
experimental studies have focused on very specific material e.g.
particles in a clay dispersion [\citet{Coussot1997},
\citet{AnceyJorrot2001}], a cement paste [\citet{Geiker2002}], a
foam [\citet{Cohenaddad2007}] or coal slurries
[\citet{sengun1989a,sengun1989b}]. This poses a problem: can we
use the results obtained in studies performed with noncolloidal
particles in clay dispersions to predict the behavior of a mortar
(i.e. particles in a cement paste)? In other words, how far can
these studies be used to predict the behavior of any kind of
noncolloidal particles suspended into any kind of yield stress
fluid? Let us review some of the results obtained in the
literature regarding the viscoelastic properties, the yield
stress, and the flow properties, in order to specify this problem.

As yield stress fluids are viscoelastic solids under the yield
stress, it is worth looking at the problem of particles in
viscoelastic materials. The influence of particles on the linear
viscoelastic properties of materials have been a little more
studied (see e.g. \citet{Poslinski1988,see2000,walberer2001}) than
their influence on yield stress fluids as they are of great
importance for the design of composite material and filled
polymers. However, it is difficult to infer a general result from
these studies: e.g. \citet{see2000} find that 40\% of spherical
80$\mu$m polyethylene particles increase the elasticity of a
silicon oil and a Separan polymer solution by a factor 5, whereas
they find that 40\% of spherical 60$\mu$m polyethylene particles
increase the elasticity of the same matrixes by a factor 8; on the
other hand, \citet{Poslinski1988} find that 40\% of 15$\mu$m glass
beads increase the elasticity of thermoplastic polymers by a
factor 15. Therefore, one may still wonder what is the influence
of the exact physicochemical origin of the viscoelasticity of the
material on the suspension mechanical properties, and if there are
specific physicochemical interactions between the large particles
and the matrix in the various studies quoted above.

Only a few studies deal with the influence of the noncolloidal
particles on the suspending fluid yield stress
[\citet{Coussot1997}, \citet{AnceyJorrot2001},
\citet{Geiker2002}]. Moreover, these studies provide extremely
different results; e.g., when spherical particles are embedded at
a volume fraction corresponding to 70\% of the maximum packing
fraction in a cement paste, \citet{Geiker2002} find that the yield
stress of the paste is increased by a factor 50, whereas
\citet{AnceyJorrot2001} find that when they are embedded in a clay
dispersion the yield stress is increased only by a factor 2. (Note
however that such discrepancy may find its origin in the yield
stress measurement method as \citet{Geiker2002} use a
Herschel-Bulkley fit of the flow curve, and \citet{chateau2007}
have shown that such a fit provides an overestimation of the yield
stress of the suspension). Moreover, \citet{AnceyJorrot2001} found
in some cases that the suspension yield stress can be lower than
the suspending paste yield stress; as pointed out by
\citet{chateau2007}, this should not occur if the noncolloidal
particle interact only mechanically with the paste. As a
consequence, \citet{AnceyJorrot2001} propose an interpretation
based on a local depletion of the colloidal particles near the
large particles. Theses results are therefore unlikely to apply if
the same particles are embedded in another paste: the experimental
results and their theoretical interpretation do not depend only on
the suspending paste yield stress but also on the composition of
the suspending paste and cannot be applied if e.g. the suspending
paste is not a colloidal suspension.

\citet{sengun1989a,sengun1989b} have investigated the flow
properties of suspensions of large particles in a colloidal
suspension: they have studied experimentally bimodal coal
slurries, and have provided a general model to predict the
viscosity of such suspensions. Their model relates the shear-rate
dependent viscosity of the material to the large particle volume
fraction and the fine particles suspension rheological properties,
independently of their exact composition. Such a model should
therefore apply to suspensions of any large particles in any
non-Newtonian fluid; however, unfortunately,
\citet{sengun1989a,sengun1989b} did not perform experiments for
non-Newtonian suspending fluids of other microscopic composition
than coal slurries. Moreover, they did not study the problem of
the influence of the large particles on the yield stress. Note
that a critical review of existing theories aiming at describing
the influence of noncolloidal particles in non-Newtonian fluids
can be found in \citet{chateau2007}.

As a consequence of these results, it is of high importance to
clarify the cases where the mechanical properties of the
suspension depend only on the mechanical properties of the
suspending fluid and on the large particle volume fraction and
size distribution; this should provide results applicable to any
particles in any yield stress fluid. Any departure from the
generic results would then be the result of specific
physicochemical interactions (or specific slippage at the
paste/particle interface) and would justify a specific study with
the particular particles and particular paste involved. With the
aim of providing such generic results, we perform an experimental
study on a broad range of materials.

In this study, we focus on the case where scale separation is
possible between the paste microstructure (i.e. the colloidal
particle size in the case of dense suspensions) and the
noncolloidal particles in suspension. Then, a simplification
occurs: these materials can be considered in a first step as
noncolloidal particles embedded in a paste (e.g. fresh concrete
$\approx$ sand and granulate + cement paste; debris-flows
$\approx$ rocks + mud). Therefore, we will consider the paste as a
continuum medium, of known mechanical properties, in which the
noncolloidal particles are embedded. Moreover, we do not want to
study a specific case (e.g. gravel in cement) but we want to
understand what happens in the general case of any rigid
noncolloidal particles embedded in any yield stress fluid, i.e. we
focus on the purely mechanical contribution of the particles to
the paste behavior, independently of the physicochemical
properties of the materials. In order to achieve this goal, we
must check several things: (i) that the particle size is much
larger than the paste microstructure size, (ii) that the results
depend only on the mechanical properties of the paste i.e. that
they are independent of the physicochemical origin of the yield
stress, (iii) that the results are independent of the noncolloidal
particles size (when the particles are monodisperse), (iv) that
there are neither particle/particle nor particle/paste
physicochemical interactions.

Finally, the problem we deal with is the homogenization of a
suspension of isotropically distributed monodisperse rigid spheres
dispersed in a continuum yield stress fluid. The present paper is
devoted to the experimental realization of this problem, while
\citet{chateau2007} deal with the theoretical problem. Here, we
focus on the behavior of the pastes in their solid regime, i.e. on
the influence of the particles on the elastic modulus and the
yield stress.

In Sec.~\ref{section_display}, we present the materials employed
and the experimental setup. We present the elastic modulus
measurements in Sec.~\ref{section_elastic}, and the yield stress
measurements in Sec.~\ref{section_yield}. We finally show in
Sec.~\ref{section_relationship} that the yield
stress/concentration relationship is related to the elastic
modulus/concentration relationship through a very simple law, in
agreement with recent results from a micromechanical analysis
presented in a companion paper [\citet{chateau2007}].

\section{Materials and methods}\label{section_display}

The materials and procedures presented hereafter are designed to
fulfill the requirements for studying the purely mechanical
contribution of an isotropic distribution of rigid monodisperse
particles to the yield stress fluids behavior.

\subsection{Pastes and particles}\label{section_materials}

As a suspending paste, we use various pastes: an emulsion, a
physical gel, and a colloidal suspension.

The emulsion is a water in oil emulsion. As the continuous phase,
we use a dodecane oil in which a Span 80 emulsifier is dispersed
at a 7\% concentration. A 300g/l CaCl$_2$ solution is dispersed in
the oil phase at 6000rpm during 1 hour. In the emulsion the origin
of the yield stress and elasticity is the surface tension between
the droplets [\citet{Larson1999}]. The microstructure scale is
thus given by the droplets which have a size of order 1$\mu$m from
microscope observations. The yield stress can be varied by varying
the droplet concentration: we vary the droplet concentration
between 70 and 90\%, and obtain 5 materials of yield stress
between 6 and 100Pa.

The physical gel is Carbopol dispersion. Here we use a Carbopol
980 (from Noveon) dispersed in water at a 0.7\% concentration,
that provides a Carbopol gel of 100Pa yield stress. The Carbopol
is dispersed at 1000rpm during 30min, then neutralized with NaOH
at pH=7. The gel is then stirred during an entire day to ensure
homogeneity of the material. The exact structure of Carbopol gels
is poorly known, and it depends a lot on the Carbopol used.
Basically, the polymers arrange in roughly spherical blobs which
tend to swell in water [\citet{ketz1988,Carnali1992}]; above a
sufficient concentration, the blobs are squeezed together and the
material develops a yield stress. The microstructure size of these
materials is badly known; from Cryo-SEM experiments
[\citet{Kim2003}] one sees that it may be as low as a few
micrometers; microrheological measurements by \citet{Oppong2006}
also indicate that it is larger than 1$\mu$m. However, a modelling
of their behavior gives an indirect indication that the blobs may
have a typical size of 100$\mu$m (it actually depends on pH). We
tried to observe the microstructure of the Carbopol gel with an
environmental SEM, but we failed to see anything; that is why, in
order to ensure scale separation between the particles and the
gel, we used particles as large as 2mm in the case of Carbopol
gels.

The colloidal suspension is a bentonite suspension; it is made of
(smectite) clay particles of length of order 1$\mu$m and thickness
10nm. Water molecules tend to penetrate between the elementary
layers composing each particle which thus swell; the particles are
then squeezed together but also interact through electrostatic
forces; it results in a yield stress. The yield stress can be
varied by varying the particle concentration: we vary the
concentration between 3 and 9\%, and obtain 6 materials of initial
yield stress between 6 and 150Pa. Note that we give the 'initial'
value of the yield stress (i.e. 100s after a preshear at high
shear rate) as these suspensions are thixotropic: their yield
stress (and elastic modulus) increases with the rest time.

Finally, we have prepared materials with 3 kinds of yield stress
physicochemical origins (jammed elastic blobs, surface tension,
colloidal interactions). If we obtain the same behavior when
adding solid particles in the 3 materials, this ensures that there
is no contribution from specific particles/material
physicochemical interactions. Another important test arises from
the bentonite suspension, which is thixotropic: if the yield
stress evolution kinetics remains unchanged when adding the
particles, then it is another indication that there are no
physicochemical interactions between the particles and the
material.

The particles are spherical monodisperse beads. We use either
polystyrene beads of density 1.05, or glass beads of density 2.5.
This allows to check that there are neither beads/beads nor
beads/paste interactions: no difference should result in the
measurements performed with different particles unless there are
interactions specific to the materials used. Note in particular
that the polystyrene beads are hydrophobic whereas the glass beads
are hydrophilic: this detail may be important when the particles
are embedded in the emulsion as the hydrophobic particles are then
preferably surrounded by the continuous phase (the oil) whereas
the hydrophilic particles are preferably surrounded by the
dispersed phase (the water). We test various particle diameters:
80, 140, 315$\mu$m in the case of the polystyrene beads, and 140,
330 and 2000$\mu$m in the case of the glass beads.

In the bentonite suspension and emulsion cases, the bead size is
much larger than the paste microstructure; in the case of the
Carbopol gel, it is undoubtedly true for the mm beads only and the
comparison of the results between the $\mu$m and mm bead will give
an indication of the relevant microstructure scale: the results
should be independent of the bead size unless this size is as low
as the microstructure scale.

We study suspensions of noncolloidal particles in yield stress
fluids at a volume fraction $\phi$ ranging between 0 and 50\%.
When preparing a suspension of beads in a yield stress fluid, the
insertion of air is unavoidable. However, methods such as
centrifugation to remove the bubbles cannot be used if we want to
ensure that the materials remain homogeneous and isotropic as
explained in Sec. \ref{section_display}\ref{section_rheom}. It is
therefore important to check that the air content is negligible.
E.g., with PS particles larger than 250$\mu$m in a bentonite
suspension, we observed that for more than 30\% of beads the air
content was of the order of 5\% of the material; this did not
occur with glass beads: the air is more likely to be entrained by
the hydrophobic (PS) particles in a suspension. It has to be noted
that the effect of air is not negligible: it changes not only the
continuous phase mechanical properties [\citet{Larson1999}] but
also the effective bead volume fraction, which is a sensitive
parameter at high volume fractions. As an example, a 40\%
suspension of 315$\mu$m PS beads in a bentonite suspension yields
an elastic modulus 30\% lower than a 40\% suspension of 330$\mu$m
glass beads in the same bentonite suspension, because of a 6\% air
content in the case of the PS beads. We chose to work with a
constant volume of material in order to check that the air content
is always lower than 1\%. All the measurements we present in this
paper were performed on materials with as negligible an air
content. We chose not to study volume fractions higher than 50\%
because we failed to prepare reproducible materials at such high
volume fractions, probably because of the unavoidable presence of
air in the materials in these cases.

Finally, note that the beads need to be carefully washed in order
to avoid surface interactions due to remaining
surfactant/stabilizer at the surface of the polystyrene beads
following their production process. The beads are washed in an
ultrasound bath during 30 minutes and then dried. We observed that
when the unwashed beads are embedded into a Carbopol gel, it
actually results in a lower yield stress than when the washed
beads are suspended, indicating residual surface effects. A single
washing is enough to ensure a reproducible state.

Note however that despite all the precautions we took, we will
show in the Appendix A that the particles may still induce non
mechanical effects that we do not understand; however, we will
present in this Appendix A a method to eliminate unambiguously
from the analysis materials in which physicochemical interactions
between the particles and paste may have occurred.

\subsection{Rheological methods}\label{section_rheom}

Most rheometric experiments are performed within a vane in cup
geometry (inner radius $R_i=12.5$mm, outer cylinder radius $R_e
=18$mm, height $H=45$mm) on a commercial rheometer (Bohlin C-VOR
200) that imposes either the torque or the rotational velocity
(with a torque feedback). In order to avoid wall slip
[\citet{Coussot2005}], we use a six-blade vane as an inner tool,
and we glue sandpaper of roughness equivalent to the size of the
particles on the outer cylinder wall. For the 2mm particles, we
use another six-blade vane in cup geometry (inner radius
$R_i=22.5$mm, outer cylinder radius $R_e =45$mm, height $H=45$mm).
Working within these wide-gap geometries allows to study easily
large particles and to ensure that for all the materials studied,
there are enough particles in the gap to consider that we measure
the properties of a continuum medium (the suspension).

We measure the elastic modulus $G'(\phi)$ and yield stress
$\tau_c(\phi)$ of the paste as a function of the volume fraction
$\phi$ of large particles embedded in the pastes. As long as we
work in the linear regime of the materials, the stress
inhomogeneities in the wide gap geometry do not affect the elastic
modulus measurements, e.g. for an elastic deformation resulting in
an angular displacement $\theta$ of the inner tool and a torque
$T$, the elastic modulus $G'$ can be computed as $\frac{T}{4\pi
H\theta}(\frac{1}{R_i^2}-\frac{1}{R_e^2})$. On the other hand, the
shear stress $\tau$ continuously decreases within the gap: the
shear stress at a radius $R$ is $\tau(R)=\frac{T}{2\pi HR^2}$.
Therefore, one has to choose a definition of the shear stress that
is measured in a given rheological experiment. Here, we want to
perform yield stress measurements; whatever the measurement method
we choose, yield first occurs where the stress is maximal i.e.
along the inner virtual cylinder. As consequence, we define the
shear stress measurement as $\tau(R_i)=\frac{T}{2\pi HR_i^2}$, so
that the yield stress $\tau_c$ is correctly measured (any other
definition would provide an underestimation). Anyway, we will
focus on the evolution of the dimensionless elastic modulus
$G'(\phi)/G'(0)$ and the dimensionless yield stress
$\tau_c(\phi)/\tau_c(0)$ with the bead volume fraction $\phi$,
which should be independent of a particular definition of $\tau$.

\subsubsection*{\it Elastic modulus measurement method}

The elastic modulus G' is determined through oscillatory shear
experiments in the linear regime: in most experiments, an
oscillatory shear stress of amplitude $\tau_0$ is applied at a
frequency of 1Hz. As we work with a controlled stress rheometer,
an oscillatory shear stress is imposed rather than an oscillatory
shear strain, in order to get accurately small deformations. The
amplitude $\tau_0$ depends on the sample: it is chosen so as to
ensure that the strain induced on the tested material is lower
than $10^{-3}$, so that all materials are tested in their linear
regime. These experiments were performed with several different
amplitudes on some materials in order to check the independence of
the results on the choice of $\tau_0$ (see
Sec.~\ref{section_elastic}, Fig.~\ref{fig5}a). We also checked the
independence of the results on the frequency (see
Sec.~\ref{section_elastic}, Fig.~\ref{fig5}b): even if the elastic
modulus $G'(0)$ of the paste may depend on the frequency, the
dimensionless modulus $G'(\phi)/G'(0)$ should not depend on it.

\begin{figure}[htbp] \begin{center}
\includegraphics[width=11.5cm]{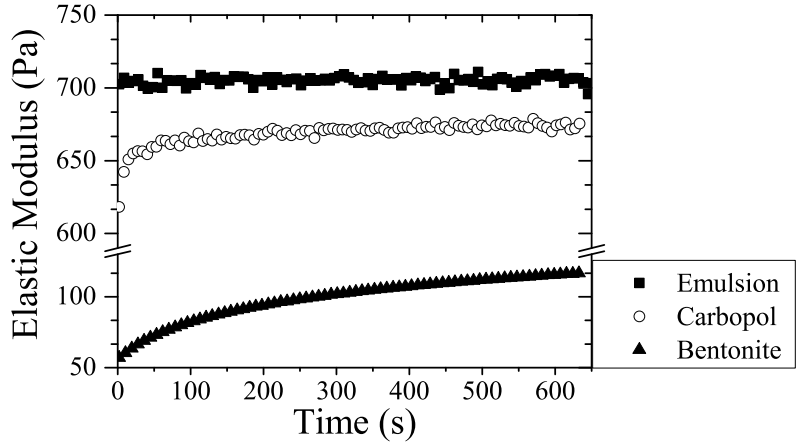}
\caption{Elastic modulus G' vs. time after loading when an
oscillatory shear stress of amplitude $\tau_0=0.3$Pa is applied at
a frequency of 1Hz to an emulsion (squares), a Carbopol gel (open
circles), and $\tau_0=0.1$Pa at 1Hz to a bentonite suspension
(triangles).}\label{fig1}
\end{center} \end{figure}

In Fig.~\ref{fig1}, we present elastic modulus measurements
performed on the 3 pastes. We observe that the elastic modulus of
the bentonite suspension strongly increases with the time of rest
(by a factor 2 in 5 minutes); this is characteristic of the aging
at rest of thixotropic materials [\cite{Coussot2006}]. On the
other hand, the emulsion elastic modulus is constant within 0.5\%
and the Carbopol gel elastic modulus increases only slightly (5\%
during 10 minutes).

\subsubsection*{\it Yield stress measurement method}

Performing a yield stress measurement is still challenging
[\citet{Coussot2005}], and there are a lot of methods designed to
evaluate this yield stress; several were used recently by
\citet{uhlherr2005} on various yield stress fluids. These methods
may give different results if e.g. the material ages with the time
of rest [\citet{Cheng1986}]. In this case the yield stress at
which a flow stops (the dynamic yield stress) differs for the
yield stress at which the flow starts (the static yield stress),
and the later is a function of the time of rest in the solid
state. The static yield stress can be measured through the vane
method [\citet{nguyen1985,Liddell1996}]: in this method the vane
tool is driven at a low velocity, and the yield stress for a given
time of rest is defined by the overshoot presented by the shear
stress (if the material is thixotropic) in a shear stress vs.
strain plot, or by the shear stress plateau if it is not
thixotropic. The static yield stress can also be measured by means
of the simple inclined plane test [\citet{Coussot1995}], by
measuring the angle between a plane and the horizontal for which a
given layer of material starts to flow along this plane under the
effect of gravity; the inclined plane can also be used to measure
the dynamic yield stress by measuring the thickness of a material
when its flow stops for a given angle of the plane. This last
property can also be measured through the slump test
[\citet{pashias1996,Roussel2005}], or through creep tests
[\citet{Coussot2006}]: if a creep stress above the yield stress is
imposed then the material flows steadily whereas for a creep
stress below the yield stress there is a creep flow that is
slowing down at any time and the strain tends to saturate. A
simple way to evaluate both the static yield stress and the
dynamic yield stress is to impose increasing and decreasing shear
stress ramps [\citet{uhlherr2005}], or increasing and decreasing
shear rate ramps.

\begin{figure}[htbp] \begin{center}
\includegraphics[width=10cm]{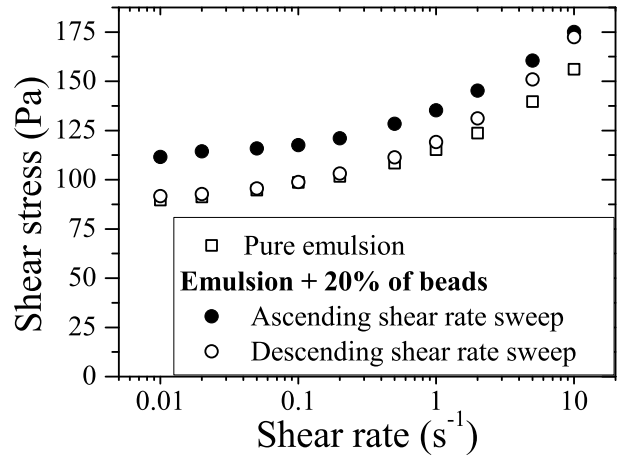}
\caption{Shear stress vs. shear-rate for ascending/descending
shear rate sweeps in a pure emulsion (open squares) and for the
same emulsion filled with 20\% of 140$\mu$m PS beads (filled/open
circles).}\label{fig2}
\end{center} \end{figure}

A question is now: are all these methods relevant in the case of
suspensions of noncolloidal particles in yield stress fluids? In
order to illustrate the problems that appear when trying to
perform a yield stress measurement on these materials, we present
an example of ascending/descending shear rate sweeps in
Fig.~\ref{fig2} for a pure emulsion, and for the same emulsion
filled with 20\% of 140$\mu$m PS beads. In these experiments,
constant shear rates increasing from 0.01 to 10s$^{-1}$ and then
decreasing from 10 to 0.01s$^{-1}$ were applied during 30s, and
the stationary shear stress was measured for each shear rate
value.

As the emulsion is not thixotropic, we observe the same curve for
the ascending/descending shear rate sweeps, and the low shear rate
value provides a good evaluation of the yield stress value (here
90Pa). We would expect to find the same independence when the
emulsion is filled with particles. Surprisingly, we find that the
shear stress during the ascending shear rate sweep differs from
the shear stress during the descending shear rate sweep. From
Fig.~\ref{fig2}, the static yield stress would then be 110Pa
whereas the dynamic yield stress would be 90Pa as in the pure
emulsion. However, we find that any measurement performed after
this experiment gives a static yield stress equal to the stopping
yield stress, i.e. 90Pa. Therefore, there have been some
irreversible change. This irreversible change is probably particle
migration towards the low shear zones; this phenomenon is well
documented in the case of suspensions in Newtonian fluids
[\citet{Leighton1987b,Abbott1991,Graham1991,Phillips1992,Nott1994,Mills1995,Morris1999,Shapley2004,Ovarlez2006}]
but is still badly known in yield stress fluids (some studies
exist however in viscoelastic fluids
[\citet{tehrani1996,huang2000,lormand2004}]). An indication that
there is indeed migration is given by the experiments performed on
suspensions of particles in a Carbopol gel: as the Carbopol gel is
transparent, we have been able to observe qualitatively the
particle distribution, and we observed that there seems to remain
not a single particle near the vane tool after shearing the
material. On the other hand, at low shear rate in a Couette
geometry the flow of yield stress fluids is localized near the
inner tool [\citet{Coussot2005}] because the shear stress
decreases continuously from the inner radius to the outer radius
and is then lower than the yield stress only in a fraction of the
gap near the outer cylinder.  As a consequence, at low shear rate
during the descending shear rate sweep, only the pure emulsion is
flowing and contributes to the shear stress as there are no more
particles near the inner tool:  this explains why the same
apparent value of the yield stress is found in the suspension
during the descending sweep as in the pure emulsion with this
method; on the other hand, the yield stress at the beginning of
the very first ascending sweep is that of the suspension as
migration has not occurred yet.

This implies that we cannot use a method based on a shear flow
such as creep tests and shear rate ramps, and that we cannot
preshear our materials. Note that this casts doubt on the
measurements performed e.g. by \citet{Coussot1997} and
\citet{Geiker2002}: their yield stress measurement is the low
shear rate limit of a flow curve; we can legitimately wonder if
their results are really characteristic of the materials at the
mean volume fraction. On the other hand, the use of the slump test
by \citet{AnceyJorrot2001} is likely to provide a fair evaluation
of the yield stress as the slump flow may have induced only a
small deformation.

\begin{figure}[htbp] \begin{center}
\includegraphics[width=11.5cm]{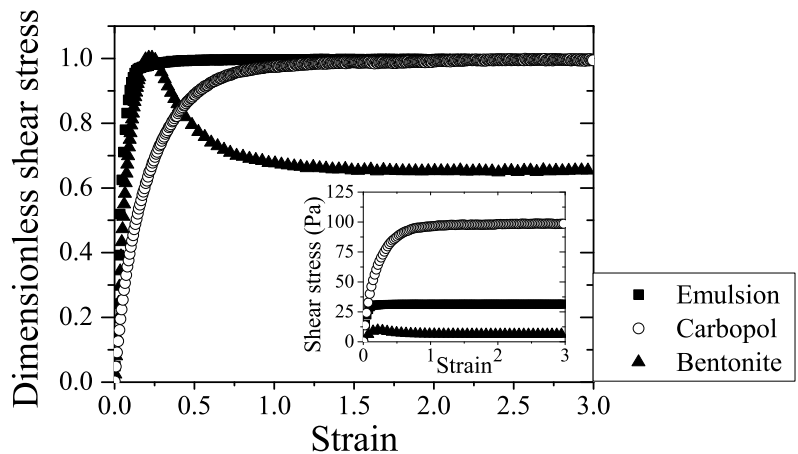}
\caption{Dimensionless shear stress vs. strain when slowly
shearing a material from rest at $10^{-2} s^{-1}$, after a 600s
rest time: emulsion (squares), Carbopol gel (empty circles), and
bentonite suspension (triangles); the shear stress is divided by
the defined yield stress. Inset: stress vs. strain for the same
experiments.}\label{fig3}
\end{center} \end{figure}

That is why for the yield stress measurements we chose to perform
a single measurement at a constant velocity on each sample,
without any initial preshear. Fig.~\ref{fig3} shows the shear
stress vs. strain for yield stress measurement experiments
performed in a bentonite suspension, an emulsion and a Carbopol
gel.

In all cases there is first a linear increase of stress with
strain: this corresponds to the elastic deformation of the
materials. In the case of the emulsion and the Carbopol gel, there
is then a plateau, corresponding to the plastic flow at low shear
rate: this defines the yield stress of these materials. In the
case of the bentonite suspension, there is an overshoot, followed
by a slow decrease of the shear stress: the peak defines the yield
stress, the decrease corresponds to destructuration of the
material under the shear flow. Note that in the case of the
bentonite suspension and the emulsion, the maximum (or the
plateau) is reached for a strain of order 0.2, whereas in the case
of the Carbopol gel one has to wait for a strain of order 1. In
this later case, it is thus likely that the particle distribution
starts to be anisotropic at the yield point (see e.g.
\cite{Gadala1980,parsi1987} for the case of particles in a
Newtonian fluid); we will comment on this point hereafter.

\subsubsection*{\it Summary of the procedure}

Finally the procedure is the following:

\begin{itemize}
\item a suspension of beads in a yield stress fluid is prepared at
a given volume fraction $\phi$ of beads ranging between 0 and 50\%
in the cylinder. There is no preshear after preparation; however,
in order to perform all the measurements in the same conditions in
the case of thixotropic materials (see the bentonite suspension
data in Fig.\ref{fig1}), i.e. to start from a same state of
destructuration of the material with and without particles, it is
necessary to strongly stir the material by hand after loading.
This procedure allows to avoid shear induced migration, and
importantly also to perform measurements on an isotropic
microstructure that allows for comparison with micromechanical
models [\citet{chateau2007}]: any controlled shear of the material
would induce an anisotropic microstructure as observed in
suspensions of particles in Newtonian fluids
[\citet{Gadala1980,parsi1987,Morris1996,Sierou2002}].

\item then, the vane tool is slowly inserted and oscillations are
performed at rest in the linear regime during 10 minutes in order
to get the elastic modulus; note that it is important to measure
the elastic modulus evolution in time (and not its mean value)
even for non thixotropic materials: this allows evidencing
undesirable physicochemical interactions between the particles and
the paste with some materials as we will show in the Appendix~A.

\item afterwards, a small rotational velocity is imposed to the
vane tool in order to get the yield stress; we checked that the
elasticity measurement is nonperturbative: the same yield stress
is measured if a zero stress is imposed instead of oscillations
before the yield stress measurement.
\end{itemize}
Finally, as the yield stress measurement may induce migration and
a microstructure anisotropy, any new measurement requires a new
sample preparation. Note that the manual preparation does not
strictly ensure that we always get the same initial state of
destructuration in the case of thixotropic materials; that is why
the yield stress measurement is performed after 10 minutes of
rest: any slight initial irreproducibility have a negligible
influence on the structure after such a time of rest.

\subsubsection*{\it Consequences of the absence of a controlled preshear}
Usually, before any measurement on a material, a preshear is
applied in order to ensure a reproducible state. In our
experiments, we could not preshear the material by rotating the
inner tool at a high velocity: we showed that in order to avoid
shear induced migration and to ensure an isotropic distribution of
the particles, it is important to stir the material by hand
without favoring any direction of shear. It is now important to
evaluate how reproducible our preparation is, and what the absence
of a controlled preshear implies on the materials behavior.

\begin{figure}[htbp] \begin{center}
\includegraphics[width=15.9cm]{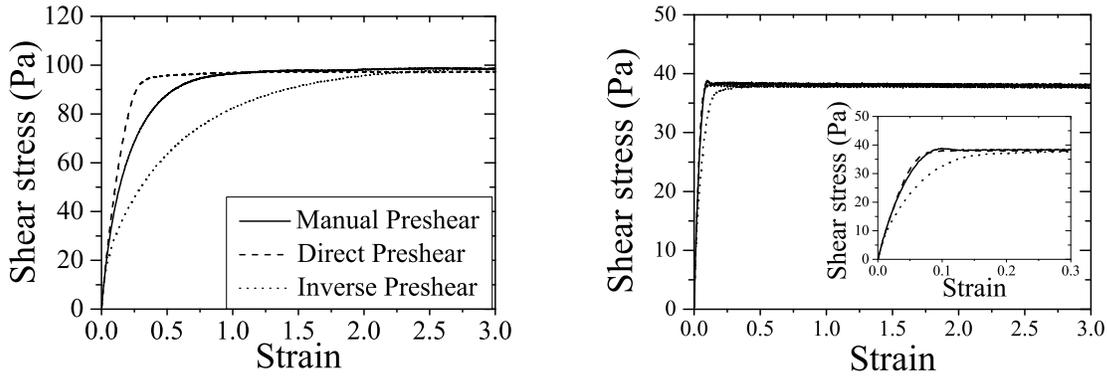}
\caption{a) Stress vs. strain when slowly shearing a pure Carbopol
gel from rest at $10^{-2}$s$^{-1}$, after a 600s rest time, in
three situations: when a 60s preshear at 10s$^{-1}$ is first
applied in the same direction as the yield stress measurement
(dashed line); when a 60s preshear at 10s$^{-1}$ is first applied
in the opposite direction to the yield stress measurement (dotted
line); when a 60s random manual preshear is first performed (solid
line). b) Same figure in the case of an emulsion; the inset is a
zoom of Fig.~\ref{fig4}b}\label{fig4}
\end{center} \end{figure}

Let us see what happens in the case of the Carbopol gel. In
Fig.~\ref{fig4}a we show the yield stress measurements performed
on the pure Carbopol gel in 3 different experiments:
\begin{itemize}
\item in a first experiment, we preshear the material by rotating
the inner cylinder at high velocity (corresponding to a shear rate
of 10s$^{-1}$) during 60s, and then we measure the yield stress by
imposing a slow velocity (corresponding to a shear rate of
0.01s$^{-1}$) in the same direction as the preshear after a 600s
rest\item in a second experiment, we preshear the material by
rotating the inner cylinder at high velocity (corresponding to a
shear rate of 10s$^{-1}$) during 60s, and then we measure the
yield stress by imposing a slow velocity (corresponding to a shear
rate of 0.01s$^{-1}$) in the opposite direction to the preshear
after a 600s rest \item in a third experiment, we preshear the
material by hand during 60s and then we measure the yield stress
by imposing a slow velocity (corresponding to a shear rate of
0.01s$^{-1}$) after a 600s rest
\end{itemize}
The role of the 600s rest before the yield stress measurement is
to erase all possible thixotropic effects.

We observe that in the three cases, the same plateau is reached.
However the critical strain for reaching this plateau varies
between 0.2 when the yield stress measurement is performed in the
same direction as the preshear, to 2 when the preshear is
performed in the opposite direction to the yield stress
measurement. This shows actually that the Carbopol gel gets
structured by the shear (note that this is not a reversible
thixotropic effect as the same state would be recovered after a
600s rest in this later case). It is also possible to show that in
the case where the direction of the preshear is opposite to the
direction of the yield stress measurement, when the stress is
unloaded from any stress before the plateau, the material does not
go back to its initial state but is irreversibly deformed; this
irreversible deformation corresponds to the progressive structural
change of the material (it is likely that the polymer chains are
progressively oriented by the shear).

Once the material is structured, the low shear rate plateau is
reached and is the same for all initial conditions: this means
that the same anisotropic state of the paste is reached, and,
importantly, that the yield stress measurement is not affected by
the preshear: whatever the reproducibility of our manual preshear,
we thus expect to get a fair evaluation of the yield stress.

If a random manual preshear is applied to the material, the
stress/strain curve falls somewhere between the two other curves
and the plateau is reached for a strain of about 1. A consequence
is that the elasticity measurement is affected: we found that the
elastic modulus measurement is reproducible within 0.5\% when a
preshear is applied with the rheometer, whereas the manual
preshear results in a reproducibility of the elastic modulus of
about 5\%: this explains partly the scatter of data in
Sec.~\ref{section_elastic}. Moreover, as mentioned above (and as
will be shown at the end of Sec.~\ref{section_yield}), in the case
of a manual preshear, the yield stress measurement is performed
for a strain sufficient for the particle distribution to be
anisotropic when the yield stress measurement is performed.

This would mean that the elastic measurement and the yield stress
measurement may not be performed exactly on the same structure of
the suspension for two reasons: (i) the interstitial Carbopol gel
is not structured the same way at rest when the elasticity
measurement is performed and at the shear stress plateau where the
yield stress is measured; (ii) the particle distribution is
isotropic for the elasticity measurement, and is possibly slightly
anisotropic for the yield stress measurement. Point (i) is
actually of no importance as far as we deal with dimensionless
quantities: as long as the interstitial paste is of a single kind
for all the elasticity measurements, and of a single kind for all
the yield stress measurements, whatever the volume fraction is,
the effect of the particles on both the dimensionless elastic
modulus and the dimensionless yield stress will be fairly
evaluated. Point (ii) is actually more of an issue and difficult
to check: a strain of order 1 is sufficient to induce a
significant anisotropy [\citet{Gadala1980}], and it may produce an
additional scatter when the yield stress data obtained on all
materials are plotted.

Finally, note that these features were also observed in the
emulsion but the phenomenon is less pronounced and the critical
strain at the yield point is at the maximum of order 0.2 i.e. the
point (ii) raised above (anisotropy of particles at the yield
point) is probably of no importance in this case. In the case of
moderately dense bentonite suspensions, there was not such
complexity as the material was initially fluid (fully
destructured). However, it is important to note that this
complexity exists as Carbopol gels are often used as a model yield
stress fluids. It seems here that they are not that model, and
that much care should be taken with these materials; another
problem with these materials when dealing with particles of
different sizes is presented in the Appendix A, .

\section{Elastic modulus}\label{section_elastic}

In this section, we summarize the results of the elastic modulus
measurements performed on all the materials. The elastic modulus
$G'$ is measured with the method presented in
Sec.~\ref{section_display} \ref{section_rheom}. We study the
evolution of the dimensionless modulus $G'(\phi)/G'(0)$ with the
volume fraction $\phi$ of noncolloidal particles for all the
materials studied.

\subsubsection*{\it Influence of the experimental parameters}

\begin{figure}[htbp] \begin{center}
\includegraphics[width=15.9cm]{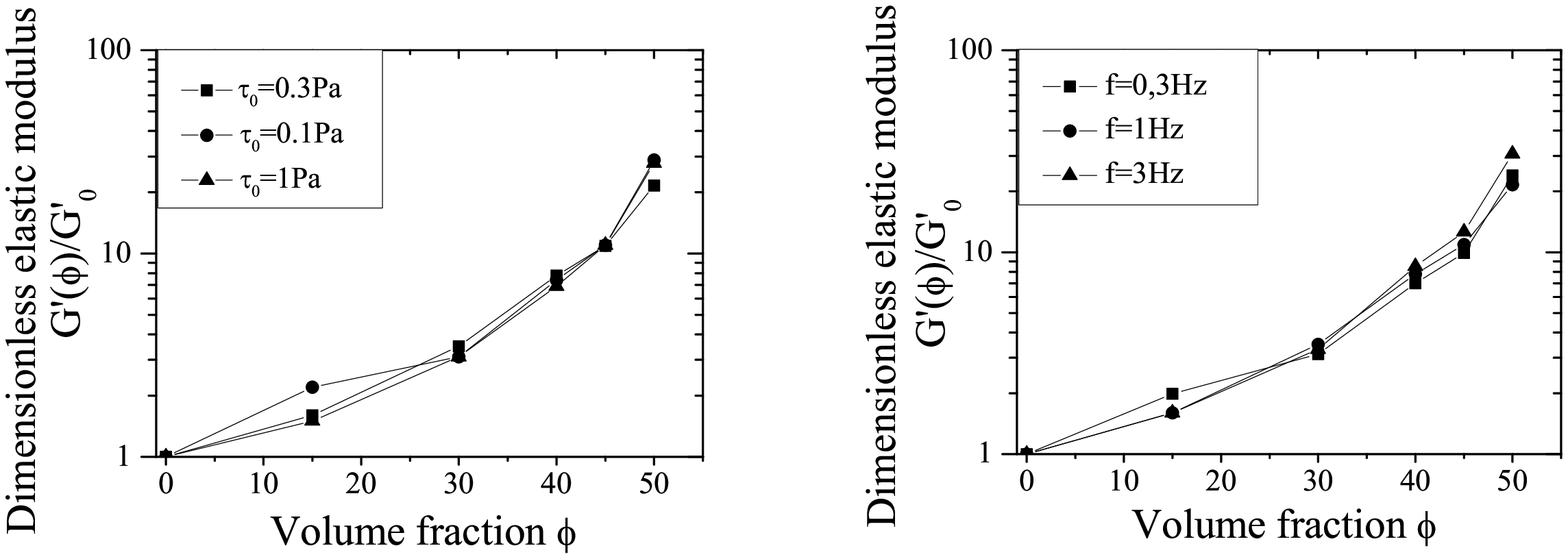}
\caption{Dimensionless elastic modulus $G'(\phi)/G'(0)$ vs. bead
volume fraction $\phi$ for a suspension of 140$\mu$m polystyrene
beads in a bentonite suspension. The dynamic measurements were
performed at a frequency of 1Hz with a stress amplitude
$\tau_0=$0.1, 0.3, and 1Pa (left figure), and also at a frequency
f=0.3, 1, and 3Hz with a stress amplitude of 0.3Pa (right
figure).}\label{fig5}
\end{center} \end{figure}

In Fig.~\ref{fig5}, we plot the dimensionless elastic modulus
$G'(\phi)/G'(0)$ vs. the beads volume fraction $\phi$ for
measurements performed at various frequencies and amplitudes on a
suspension of 140$\mu$m polystyrene beads in a bentonite
suspension. We observe that the evolution of the dimensionless
elastic modulus with $\phi$ is independent of the frequency and
the amplitude as long as we are in the linear regime: this
measurement method thus provides a fair evaluation of the
influence of the inclusion of rigid particles on the elasticity of
a yield stress fluid. We also checked that the dimensionless
elastic modulus of a suspension of 140$\mu$m polystyrene beads in
an emulsion is independent of the frequency and the stress
amplitude. In all the other experiments presented hereafter, all
the results were obtained at a frequency of 1Hz and a given stress
amplitude (dependent on the material studied) in the linear
regime.

\subsubsection*{\it Structuration kinetics}

\begin{figure}[htbp] \begin{center}
\includegraphics[width=15.9cm]{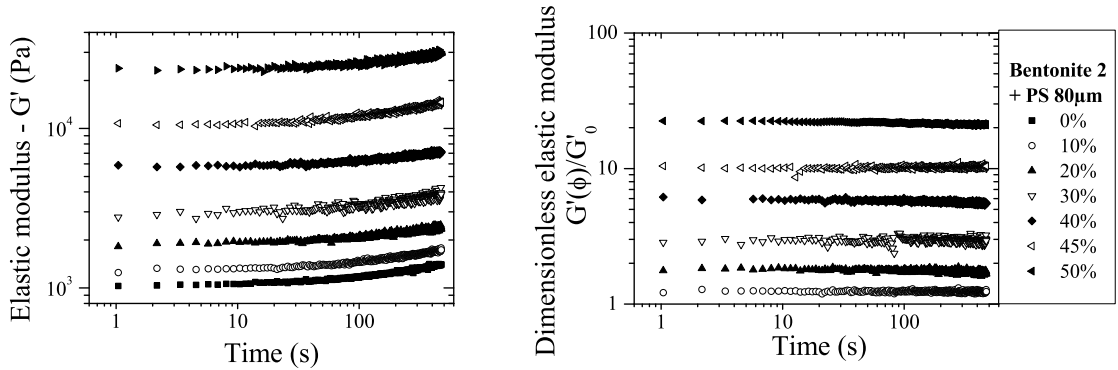}
\caption{a) Elastic modulus $G'(\phi,t)$ vs. rest time for
suspensions of 80$\mu$m polystyrene beads in a bentonite
suspension, for bead volume fraction $\phi$ ranging between 0 and
50\% (see legend). b) Dimensionless elastic modulus
$G'(\phi,t)/G'(0,t)$ vs. rest time for the same
materials.}\label{fig6}
\end{center} \end{figure}

In Fig.~\ref{fig6}a, we plot the evolution with the rest time $t$
of the elastic modulus $G'(\phi,t)$ of suspensions of 140$\mu$m
polystyrene beads in a bentonite suspension for a volume fraction
of beads $\phi$ ranging between 0 and 50\%. We plot the
dimensionless elastic modulus $G'(\phi,t)/G'(0,t)$ as a function
of time for the same materials in Fig.~\ref{fig6}b.

We observe in Fig.~\ref{fig6}a that the elastic modulus is an
increasing function of the time of rest whatever the bead volume
fraction; this evolution is due to the bentonite suspension
structuration at rest. We also observe in Fig.~\ref{fig6}b that
the dimensionless elastic modulus $G'(\phi,t)/G'(0,t)$ is
independent of time. This means that the effect of time and of the
volume fraction can be separated, i.e. the elastic modulus of the
suspension at volume fraction $\phi$ can be written as
$G'(\phi,t)=G'(0,t)f(\phi)$. In other words, this means that the
structuration kinetics of the bentonite suspension is not affected
by the presence of beads: this is consistent with our aim that
there should be no physicochemical interactions between particles
and paste. Finally, the function $f(\phi)$ is what we are seeking
i.e. the function accounting for the mechanical strengthening of
the material due to the presence of rigid inclusions.

We show in the Appendix A that even in the case of non (or
slightly) thixotropic materials, it is worth measuring the elastic
modulus evolution in time with and without particles: it is a
powerful mean for evidencing specific physicochemical interactions
between the particles and some materials.

\subsubsection*{\it Influence of the materials, the bead size and the paste elastic modulus (and yield stress) value}

\begin{figure}[htbp] \begin{center}
\includegraphics[width=12cm]{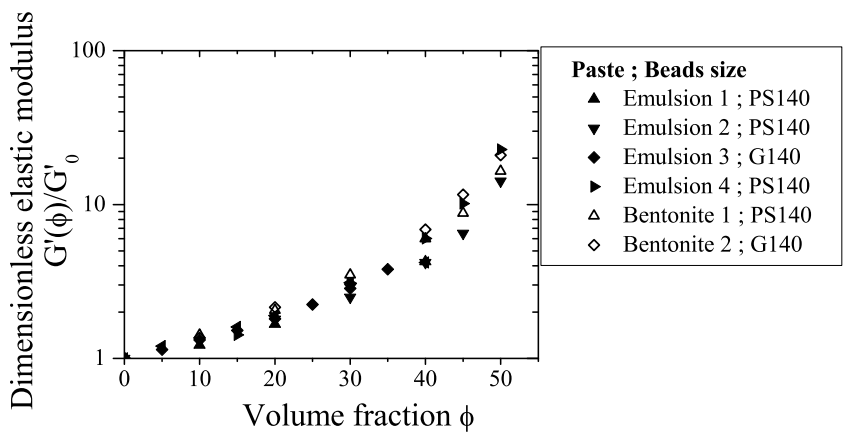}
\caption{Dimensionless elastic modulus $G'(\phi)/G'(0)$ vs. the
beads volume fraction $\phi$ for a suspension of 140$\mu$m
polystyrene (PS) and glass (G) beads in 2 different bentonite
suspensions and 4 different emulsions (of elastic modulus varying
between 300Pa and 3000Pa).}\label{fig7}
\end{center} \end{figure}

In Fig.~\ref{fig7}, we plot the dimensionless elastic modulus
$G'(\phi)/G'(0)$ vs. the beads volume fraction $\phi$ for
suspensions of 140$\mu$m polystyrene and glass beads in two
different bentonite suspensions, and various emulsions of elastic
modulus varying between 300Pa and 3000Pa. We observe that the
evolution of the dimensionless elastic modulus is independent of
the material, i.e. it is independent of the physicochemical origin
of the material elasticity and the bead material; this shows that
there are probably no physicochemical interactions between the
particles and the pastes. Moreover, we observe that, as expected,
the results are independent of the paste elastic modulus, i.e. the
elastic modulus of the suspension at volume fraction $\phi$ can be
written as $G'(\phi)=G'(0)f(\phi)$.

\begin{figure}[htbp] \begin{center}
\includegraphics[width=11.5cm]{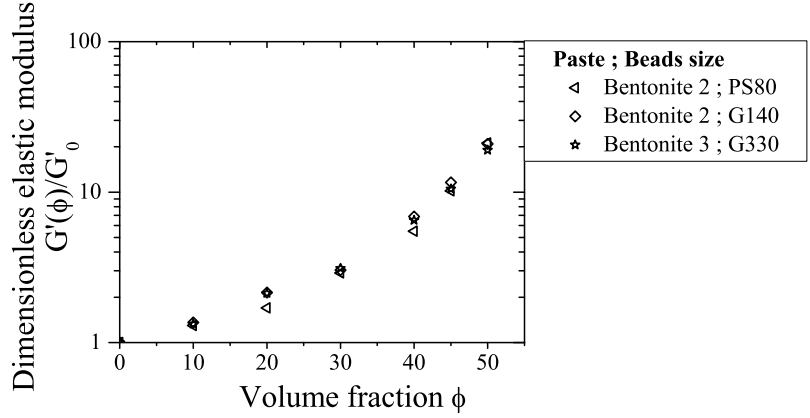}
\caption{Dimensionless elastic modulus $G'(\phi)/G'(0)$ vs. the
beads volume fraction $\phi$ for suspensions of 80, 140, and
330$\mu$m beads in a bentonite suspension.}\label{fig8}
\end{center} \end{figure}

In Fig.~\ref{fig8}, we plot the dimensionless elastic modulus
$G'(\phi)/G'(0)$ vs. the beads volume fraction $\phi$ for
suspensions of beads of various sizes in a bentonite suspension.
We observe that the evolution of the dimensionless elastic modulus
is independent of the bead size (and of the bead material, see
Fig.~\ref{fig7}); this is consistent with the absence of
physicochemical interactions between the particles and between the
particles and the pastes.

\ \\

\subsubsection*{\it Summary of the results on all materials}

\begin{figure}[htbp] \begin{center}
\includegraphics[width=15.9cm]{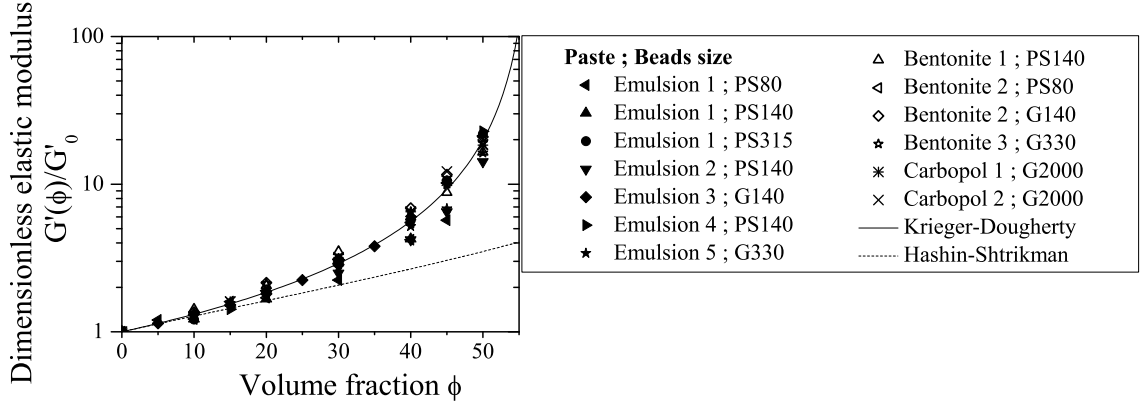}
\caption{Dimensionless elastic modulus $G'(\phi)/G'(0)$ vs. the
beads volume fraction $\phi$ for suspensions of 80, 140, and
315$\mu$m polystyrene beads (PS) and $140\mu$m, $330\mu$m and 2mm
glass beads (G) in various bentonite suspensions, emulsions and
Carbopol gels. The solid line is a Krieger-Dougherty law
$(1-\phi/\phi_m)^{-2.5\phi_m}$ with $\phi_m=0.57$. The dotted line
is the Hashin-Shtrikman bound $(2+3\phi)/(2-2\phi)$.}\label{fig9}
\end{center} \end{figure}

In Fig.~\ref{fig9}, we plot a summary of all the dimensionless
elastic modulus measurements $G'(\phi)/G'(0)$. In
Tab.\ref{tab_elastic} we summarize the mean elastic modulus values
measured on all the materials (when several measurements were
performed at a given concentration), their standard deviation, and
the number of materials on which these values were obtained.

\begin{table}[htbp]\begin{center}\begin{tabular}{|c|c|c|c|} \hline Concentration (\%) & Mean
Dimensionless & Standard Deviation (\%)
& Number of materials\\
&Elastic Modulus &&\\
\hline10 & 1.30 & 5.4 & 9\\
15 & 1.55 & 4.6 & 6 \\
20 & 1.90 & 9.3 & 9\\
30 & 2.96 & 11 & 13\\
40 & 5.62 & 17 & 12\\
45 & 9.42 & 23 & 11\\
50 & 18.9 & 14 & 10\\
\hline \end{tabular}\caption{Mean elastic modulus values and
standard deviation (in \%) as a function of the particle volume
fraction.}\label{tab_elastic}\end{center}\end{table}

We observe that we have managed to obtain the purely mechanical
contribution of the particles to the paste behavior, independently
of the physicochemical properties of the materials, i.e. the
strengthening of the materials is given by a function $f(\phi)$ of
the volume fraction $\phi$ only: the elastic modulus of an
isotropic suspension of monodisperse noncolloidal spherical rigid
particles of any size (larger than the yield stress fluid
microstructure) embedded at a volume fraction $\phi$ in a yield
stress fluid of elastic modulus $G'(0)$ (that may be a function of
time) can be written as $G'(\phi)=G'(0)f(\phi)$. Consistently, we
note that the data points all fall above the Hashin-Shtrikman
lower bound [\citet{Hashin1963}]
\bea\frac{G'(\phi)}{G'(0)}>\frac{2+3\phi}{2-2\phi}\eea which is a
theoretical bound computed in the more general case of a biphasic
material (an infinitely rigid phase embedded in a linear elastic
phase) isotropic both at the microscopic and the macroscopic
scales, with no physicochemical interactions. Note finally that
the consistency between all results, as well as their perfect
agreement with the Einstein law $G'(\phi)/G'(0)=1+2.5\phi$ at low
volume fraction, show that there is no slippage at the paste/bead
interface during the elastic modulus measurement (in the case of
full slippage, the exact result in the dilute limit would be
$G'(\phi)/G'(0)=1+\phi$ [\citet{Larson1999}]).

The effect of the particle concentration on the elastic modulus is
rather important, even for low bead volume fraction: we find
$G'(\phi)\approx 1.9\times G'(0)$ for $\phi=20\%$ and
$G'(\phi)\approx 18\times G'(0)$ for $\phi=50\%$. Note that the
scatter of data is much higher than what can be found (by the same
experimentalist i.e. with exactly the same methods) when measuring
the viscosity of suspensions of noncolloidal particles suspended
in various Newtonian fluids (see e.g. Fig.~9 in
\citet{Zarraga2000}): this is expected as flow properties
correspond to a mean over all the configurations explored during
flow, whereas the static properties are measured over a particular
configuration. Moreover, we have shown in
Sec.~\ref{section_display} that the necessary absence of a
controlled preshear is an important source of scatter. Finally, as
mentioned in Sec.~\ref{section_display}\ref{section_materials}, we
control the air content only up to 1\%; this induces a 1\%
uncertainty on the bead volume fraction. For a 50\% beads volume
fraction, from Eq.~\ref{eq_elasticity}, this yields a 10\%
uncertainty on the elastic modulus.

We observe that our data are very well fitted to a
Krieger-Dougherty law [\citet{Krieger1959}]

\bea
G'(\phi)=G'(0)\frac{1}{(1-\phi/\phi_m)^{2.5\phi_m}}\label{eq_elasticity}
\eea We find $\phi_m=0.57$ with a least-squares fit of all the
experimental data. The Krieger-Dougherty law is usually found to
fit most viscosity data of suspension of noncolloidal particles in
Newtonian fluids. Therefore, the good agreement with our data is
not surprising as the problem of the elasticity of a suspension of
rigid particles in a linear elastic material is formally similar
to the problem of the viscosity a suspension of rigid particles in
a Newtonian (thus linear) material. Note however that the
$\phi_m=0.57$ value we find is lower than the 0.63 value often
found for the viscosity of suspensions [\citet{Larson1999}] and
than the 0.605 measured locally recently by \citet{Ovarlez2006}
through MRI techniques. This discrepancy is likely due to the fact
that most experiments in the literature on the viscosity of
suspensions are performed on an anisotropic microstructure induced
by the flow [\citet{parsi1987}], whereas our experiments are
performed on an isotropic microstructure.

\section{Yield stress}\label{section_yield}

In this section, we summarize the results of the yield stress
measurements performed on all the materials. The yield stress
$\tau_c$ is measured with the method presented in
Sec.~\ref{section_display}\ref{section_rheom}. We study the
evolution of the dimensionless yield stress
$\tau_c(\phi)/\tau_c(0)$ with the volume fraction $\phi$ of
noncolloidal particles for all the materials studied.

\subsubsection*{\it Influence of the experimental parameters}

\begin{figure}[htbp] \begin{center}
\includegraphics[width=7.2cm]{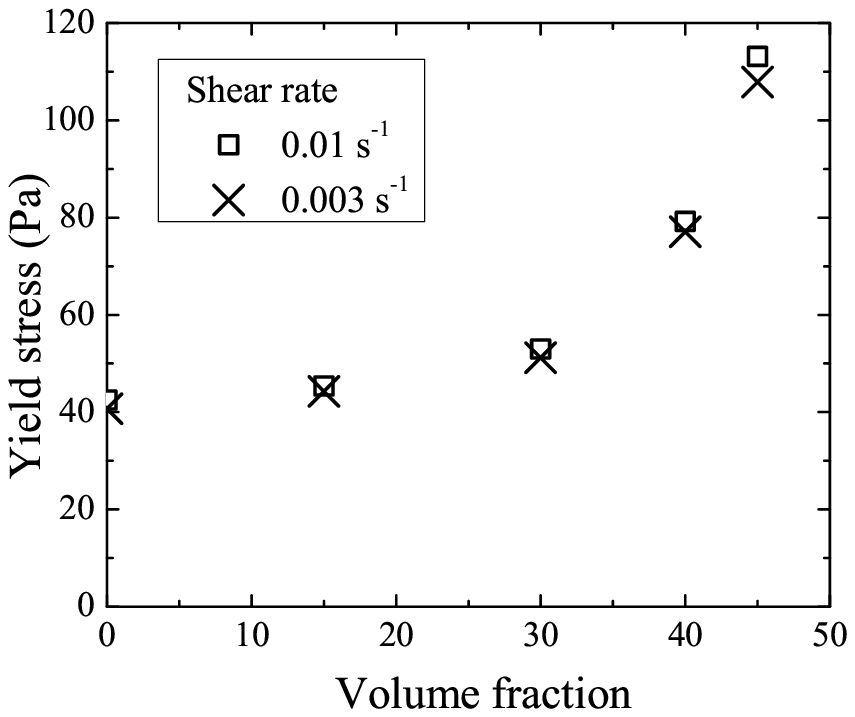}
\includegraphics[width=7.5cm]{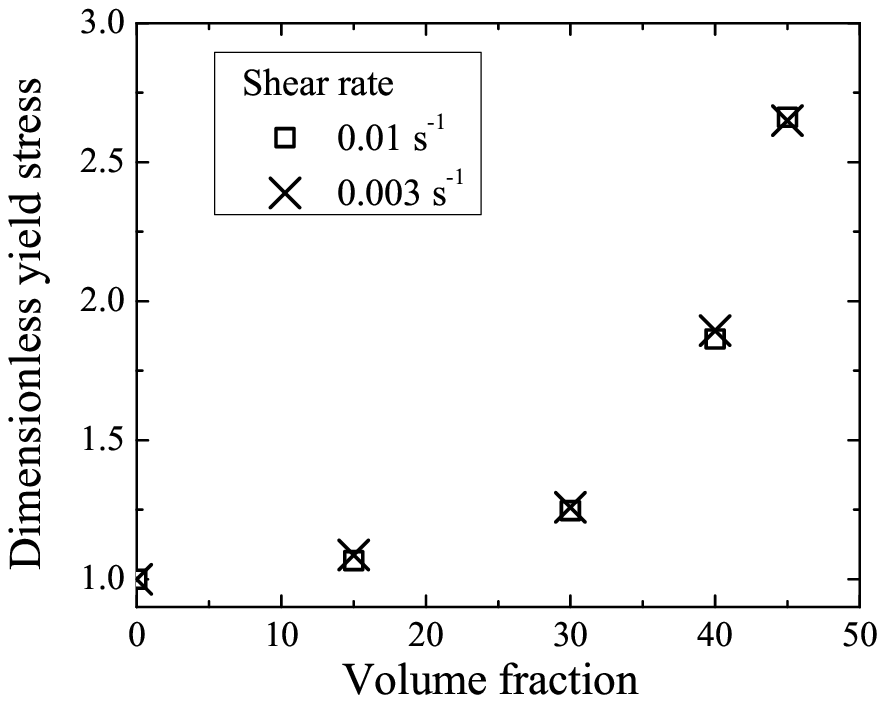}
\caption{Yield stress $\tau_c(\phi)$ (left) and dimensionless
yield stress $\tau_c(\phi)/\tau_c(0)$ (right) vs. the beads volume
fraction $\phi$ for a suspension of 140$\mu$m polystyrene beads in
an emulsion. The yield stress measurements were performed at shear
rates $\gdot=0.003s^{-1}$ (crosses) and 0.01$s^{-1}$ (empty
squares).}\label{yield_stress_velocity}
\end{center} \end{figure}

In Fig.~\ref{yield_stress_velocity}, we plot the dimensionless
yield stress $\tau_c(\phi)/\tau_c(0)$ vs. the beads volume
fraction $\phi$ for measurements performed at two different shear
rates on suspensions of 140$\mu$m polystyrene beads in an
emulsion. Importantly, we observe that even if the absolute value
of the yield stress may depend on the shear rate (and more
generally on the yield stress measurement method) the evolution of
the dimensionless yield stress with the volume fraction is
independent of the shear rate (as long as it is low enough): while
there is a 10\% difference between yield stresses measured at 0.01
and 0.003 s$^{-1}$, the dimensionless stresses values are equal
within less than 1\%. Therefore, this measurement method provides
a fair evaluation of the influence of the inclusion of rigid
particles on the resistance of a yield stress fluid. In all the
experiments presented hereafter, all the results were obtained at
a shear rate $\gdot=0.01s^{-1}$.

\subsubsection*{\it Summary of the results on all materials}

In the case of the yield stress measurements, we showed the same
features as for the elastic modulus measurements, i.e. all the
results are consistent with our aim that there should be no
physicochemical interactions between the particles and the pastes:
\begin{itemize}
\item the dimensionless yield stress $\tau_c(\phi,t)/\tau_c(0,t)$
for suspensions in a thixotropic material (the bentonite
suspension) is independent of time, i.e. the yield stress of the
suspension at volume fraction $\phi$ can be written as
$\tau_c(\phi,t)=\tau_c(0,t)g(\phi)$. This shows again that the
structuration kinetics of the bentonite suspension is not affected
by the presence of beads.

\item the dimensionless yield stress $\tau_c(\phi)/\tau_c(0)$ is
independent of the physicochemical origin of the material yield
stress, of the bead material and of the bead size.

\item the dimensionless yield stress $\tau_c(\phi)/\tau_c(0)$ is
independent of the paste yield stress.
\end{itemize}

\begin{figure}[htbp] \begin{center}
\includegraphics[width=15.9cm]{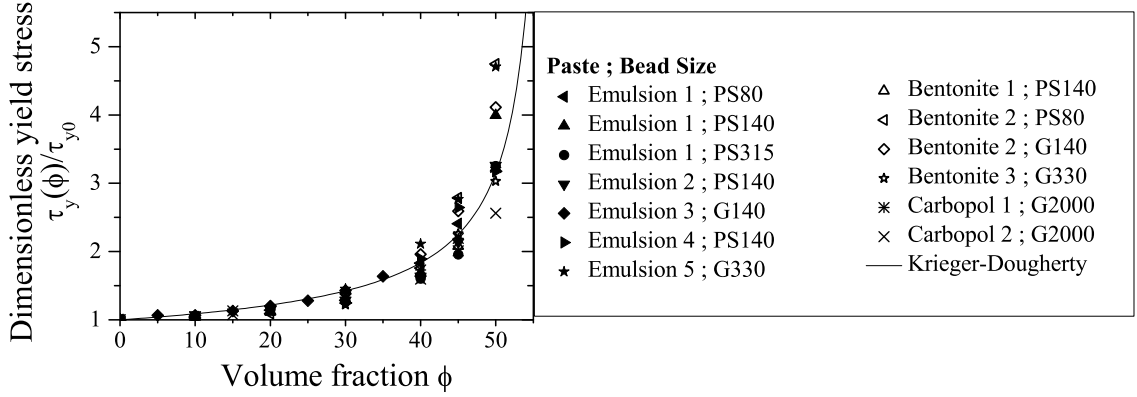}
\caption{Dimensionless yield stress $\tau_c(\phi)/\tau_c(0)$ vs.
the beads volume fraction $\phi$ for suspensions of 80, 140, and
315$\mu$m polystyrene beads (PS) and $140\mu$m, $330\mu$m and 2mm
glass beads (G) in various bentonite suspensions, emulsions and
Carbopol gels. The solid line is a law
$\sqrt{(1-\phi)\times(1-\phi/\phi_m)^{-2.5\phi_m}}$ with
$\phi_m=0.57$.}\label{fig13}
\end{center} \end{figure}

In Fig.~\ref{fig13}, we plot a summary of the dimensionless yield
stress measurements $\tau_c(\phi)/\tau_c(0)$ performed on all the
materials. In Tab.\ref{tab_yield} we summarize the mean yield
stress values measured on all the materials (when several
measurements were performed at a given concentration), their
standard deviation, and the number of materials on which these
values were obtained.

\begin{table}[htbp] \begin{center}\begin{tabular}{|c|c|c|c|} \hline Concentration (\%) &
Mean dimensionless & Standard Deviation (\%)
& Number of materials\\
&yield stress&&\\
\hline 10 & 1.05 & 1.5 & 9\\
15 & 1.12 & 2.5 & 4 \\
20 & 1.15 & 3.1 & 9\\
30 & 1.32 & 5.4 & 12\\
40 & 1.76 & 9.2 & 12\\
45 & 2.35 & 13 & 11\\
50 & 3.57 & 20 & 11\\
\hline \end{tabular}\caption{Mean yield stress values and standard
deviation (in \%) as a function of the particle volume
fraction.}\label{tab_yield}\end{center}\end{table}

We observe that we have managed to obtain the purely mechanical
contribution of the particles to the paste behavior, independently
of the physicochemical properties of the materials, i.e. the
strengthening of the materials is given by a function $g(\phi)$ of
the volume fraction $\phi$ only: the yield stress of a suspension
of monodisperse noncolloidal spherical rigid particle of any size
(larger than the yield stress fluid microstructure) embedded at a
volume fraction $\phi$ in a yield stress fluid of yield stress
$\tau_c(0)$ (that may depend on time) can be written as
$\tau_c(\phi)=\tau_c(0)g(\phi)$. As in the case of the elastic
modulus measurements, we observe a good consistency between all
results, which means that there is no (or negligible) slippage at
the paste/bead interface during the yield stress measurement.

We note that the function $g(\phi)$ is different from the function
$f(\phi)$ obtained for the dimensionless elastic modulus
measurement: the effect of the volume fraction of noncolloidal
particles on the yield stress is actually much less important than
its effect on the elastic modulus. The yield stress $\tau_c(\phi)$
at a volume fraction $\phi=30\%$ is only 40\% higher than the
yield stress $\tau_c(0)$ of the yield stress fluid, whereas its
elasticity $G'(\phi)$ is almost 200\% higher than the yield stress
fluid elasticity $G'(0)$; for $\phi=50\%$, $\tau_c(\phi)\approx
3.2\times\tau_c(0)$ while $G'(\phi)\approx 18\times G'(0)$. This
is very different from the behavior of colloidal suspension: their
yield stress follows roughly the same evolution as their elastic
modulus with the volume fraction $\psi$ of colloidal particles
(see Appendix B for an example). Finally, note that the scatter of
the yield stress values is of the order of the scatter observed in
the elastic modulus measurements.

The data are well fitted to a law \bea
\frac{\tau_c(\phi)}{\tau_c(0)}=\sqrt{\frac{1-\phi}{(1-\phi/\phi_m)^{2.5\phi_m}}}
\eea with $\phi_m=0.57$ (see Fig.~\ref{fig13}); we will provide a
justification for this law in Sec.~\ref{section_relationship}.

Finally, our observation that the yield stress of suspensions of
noncolloidal particles embedded in a thixotropic paste of
time-dependent yield stress $\tau_c(0,t)$ reads
$\tau_c(\phi,t)=\tau_c(0,t)g(\phi)$ has an interesting consequence
for materials formulation. This means that it is sufficient to
know how the interstitial paste evolves to predict the suspension
evolution at rest. This may be of importance for materials like
concretes as their behavior is hard to measure: a good knowledge
of the cement paste structuration at rest is sufficient.

\subsubsection*{\it The yield stress measurement: a destructive measurement}

\begin{figure}[htbp] \begin{center}
\includegraphics[width=9.5cm]{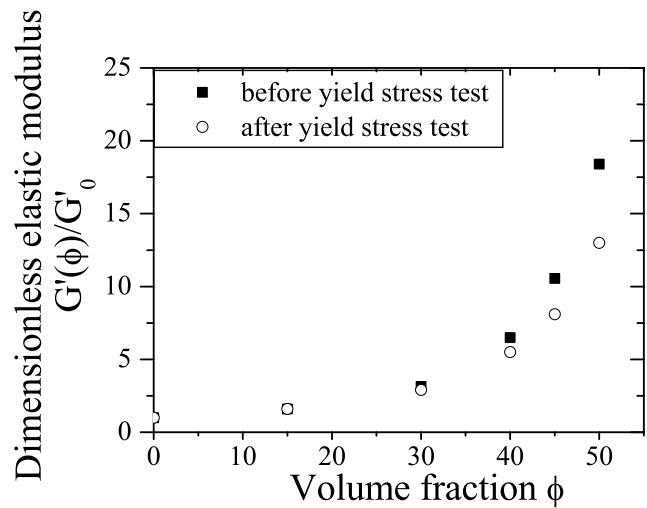}
\caption{Dimensionless elastic modulus $G'(\phi)/G'(0)$ vs. the
beads volume fraction $\phi$ for suspensions of 2mm glass beads in
a Carbopol gel before (filled squares) and after (open circles)
the yield stress
measurements.}\label{mesure_destructrice_sur_gprime}
\end{center} \end{figure}

In order to evidence the change of the material due to the yield
stress measurement, we have performed a second elasticity
measurement just after the yield stress measurement on some
pastes. In Fig.~\ref{mesure_destructrice_sur_gprime} we plot the
evolution with the bead volume fraction $\phi$ of the
dimensionless elastic modulus $G'(\phi)/G'(0)$ measured before and
after the yield stress measurement in suspensions of 2mm glass
beads in a Carbopol gel, both measurements ($G'(\phi)$ and
$G'(0)$) being performed in the same conditions (i.e. both before
or both after the yield stress measurement). In all the yield
stress measurement experiments, the inner tool was driven at a
velocity of 0.01$s^{-1}$ during 300s, resulting in a strain value
of 3.

We observe in Fig.~\ref{mesure_destructrice_sur_gprime} that the
material was changed by the slow shear: it appears to be less
stiff after the shear than before the yield stress measurement.
The change can be rather important as the dimensionless modulus
measured before the slow shear is 40\% higher than the modulus
measured after the slow shear in the case of the 50\% suspension.

Let us now examine the possible reasons for this change. First,
one should notice that as both measurements ($G'(\phi)$ and
$G'(0)$) are performed in the same conditions, the difference in
the $G'(\phi)/G'(0)$ values cannot be attributed to the small
rejuvenation of the suspending fluid by the shear, which should be
roughly the same with and without beads (note moreover from
Fig.~\ref{fig1} that the possible variation of the elastic modulus
of the Carbopol gel due to shear is of order 5\%).

A possibility, as pointed out in Sec.~\ref{section_display} is
that shear-induced migration of the particles has begun to occur
as such a phenomenon would lead to an apparent decrease in the
elastic modulus. However, it should be noted that one would not
expect a deformation of 3 to generate significant migration;
nevertheless, at such low velocity the flow is likely to be
localized near the inner cylinder and it is still possible that
the depletion of particles near the inner tool is thus a rapid
phenomenon that leads to a decrease in the apparent elastic
modulus.

Another possibility is that the suspension microstructure has
become anisotropic: it is known that under shear suspensions of
particles develop an anisotropic microstructure
[\citet{parsi1987,Morris1996}]. It has been shown by
\citet{Gadala1980}, \citet{Narumi2002} and \citet{Narumi2005} that
the stationary anisotropic structure is reached for a deformation
of order 2. It is thus highly probable that the decrease in the
elastic modulus we observe at high concentrations is a signature
of the development of a particle anisotropy due to shear.

To conclude, since we want to deal with homogeneous and isotropic
materials, only the first elasticity and yield stress measurements
can be trusted. Any other measurement requires a new preparation
of the material.

\section{Elastic modulus vs. yield stress: comparison with a micromechanical
approach}\label{section_relationship}

Proposing a theoretical value for the dimensionless elastic
modulus and the dimensionless yield stress is challenging; e.g.,
in a micromechanical approach, the final result depends on the
scheme that is chosen. However, it is shown in a companion paper
by \citet{chateau2007} that it is possible to give a general
relationship between the linear response of the materials (e.g.
its dimensionless elastic modulus $G'(\phi)/G'(0)$ as in our
study) and the dimensionless yield stress $\tau_c(\phi)/\tau_c(0)$
of a suspension of rigid particles in a yield stress fluid that is
true whatever the scheme (the way the phases interact one with the
other) as long as the particle distribution is isotropic, and
provided the strain heterogeneities are weak (i.e. this should not
be true at high volume fractions; this last point is discussed in
detail by \citet{chateau2007}).

\citet{chateau2007} find \bea
\frac{\tau_c(\phi)}{\tau_c(0)}=\sqrt{(1-\phi)\frac{G'(\phi)}{G'(0)}}\label{eq_elasticity_vs_yield_stress}
\eea Note that the micromechanical estimate of \citet{chateau2007}
is based on the following hypotheses: the particles are rigid,
monodisperse and noncolloidal; there are no physicochemical
interactions between the particles and the paste; the distribution
of the particles is isotropic. This is what we have managed to
perform experimentally, therefore, our experiments are fitted to
provide a test of these theoretical predictions.

\begin{figure}[htbp] \begin{center}
\includegraphics[width=15.9cm]{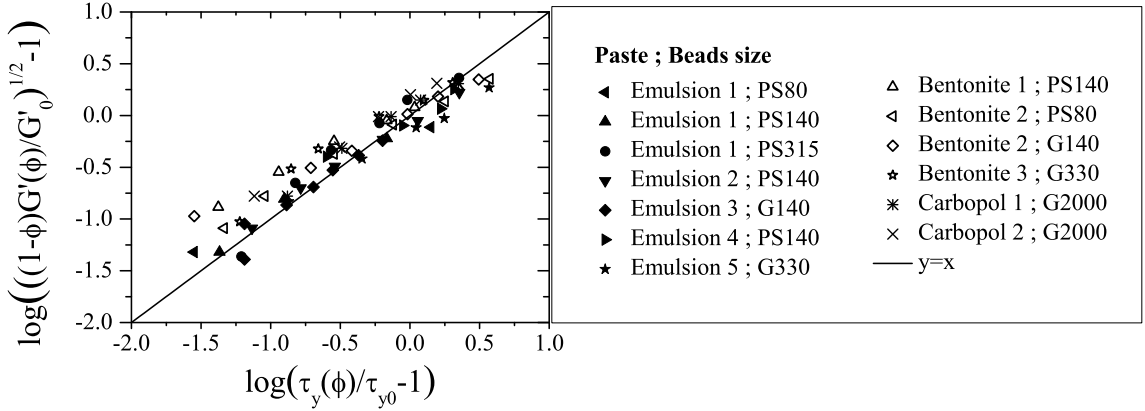}
\caption{Dimensionless yield stress
$\log\bigl(\tau_c(\phi)/\tau_c(0)-1\bigr)$ vs. some function of
the dimensionless elastic modulus
$\log\bigl(\sqrt{(1-\phi)G'(\phi)/G'(0)}-1\bigr)$ for all the
systems studied. The line is a $y=x$ plot corresponding to the
theoretical prediction of \citet{chateau2007}.}\label{fig15}
\end{center} \end{figure}

In Fig.~\ref{fig15}, we plot the dimensionless yield stress
$\tau_c(\phi)/\tau_c(0)$ vs. a function of the dimensionless
elastic modulus $\sqrt{(1-\phi)G'(\phi)/G'(0)}$ for all the
systems studied in logarithmic coordinates.

We observe a remarkable agreement between our results and the
micromechanical estimation of Eq.
\ref{eq_elasticity_vs_yield_stress} (that is plotted as a straight
line $y=x$ in these coordinates). As a consequence, combining Eq.
\ref{eq_elasticity} and Eq. \ref{eq_elasticity_vs_yield_stress},
this yields for the yield stress \bea
\frac{\tau_c(\phi)}{\tau_c(0)}=\sqrt{\frac{1-\phi}{(1-\phi/\phi_m)^{2.5\phi_m}}}\label{eq_yield_stress}
\eea with $\phi_m=0.57$, which is plotted in Fig.~\ref{fig13}, and
of course is well fitted to the experimental data.

Eq. \ref{eq_elasticity_vs_yield_stress} also means that the
critical yield strain $\gamma_c$, defined as the strain at which a
flow starts, follows a law \bea
\gamma_c(\phi)\approx\frac{\tau_c(\phi)}{G'(\phi)}=\gamma_c(0)\sqrt{(1-\phi)G'(0)/G'(\phi)}\label{eq_yield_strain}
\eea which, together with Eq. \ref{eq_elasticity} yields \bea
\gamma_c(\phi)=\gamma_c(0)\sqrt{(1-\phi)(1-\phi/\phi_m)^{2.5\phi_m}}\label{eq_yield_strain}
\eea

From Eq.~\ref{eq_elasticity_vs_yield_stress}, we see that
measuring the linear properties of suspensions helps to find the
nonlinear properties of suspensions in yield stress fluids;
theoretically, this linear response could be measured as the
viscosity of suspensions of noncolloidal particles in Newtonian
fluids. However, it should be noted that usually the viscosity
measurements of suspensions are performed on an anisotropic
structure induced by the flow [\citet{parsi1987}]. As a
consequence, most viscosity laws one can find in the literature
cannot be used to predict the yield stress of isotropic
suspensions of noncolloidal particles in yield stress fluids,
whereas our elasticity measurements can.

Finally, note that Eq.~\ref{eq_elasticity_vs_yield_stress} is very
different from what can be observed in colloidal suspension in
which the elastic modulus and the yield stress follow roughly the
same evolution with the volume fraction $\psi$ of colloidal
particles (see Appendix B for an example).

\section{Conclusion}

We have studied experimentally the behavior of isotropic
suspensions of monodisperse rigid spherical noncolloidal particles
in yield stress fluids. In order to evaluate the purely mechanical
contribution of the particles to the paste behavior, independently
of the physicochemical properties of the materials, we have
suspended beads of various sizes, made of various materials in
very different pastes (an emulsion, a microgel, and a colloidal
suspension) whose common point is to exhibit a yield stress, and
we sought consistency between the results. We focused on the
influence of the particles on the elastic modulus and the yield
stress; we used experimental procedures designed to ensure
isotropy of the suspensions. We showed that the dimensionless
elastic modulus $G'(\phi)/G'(0)$ and the dimensionless yield
stress $\tau_c(\phi)/\tau_c(0)$ depend on the bead volume fraction
$\phi$ only. We found that the elastic modulus/concentration
relationship is well fitted to a Krieger-Dougherty law
$(1-\phi/\phi_m)^{-2.5\phi_m}$ with $\phi_m=0.57$, and showed that
the yield stress/concentration relationship is related to the
elastic modulus/concentration relationship through a very simple
law $\tau_c(\phi)/\tau_c(0)=\sqrt{(1-\phi)G'(\phi)/G'(0)}$, in
agreement with recent results from a micromechanical analysis. We
now plan to study the case of bidisperse systems and of
anisotropic particle distributions. We also intend to study the
influence of the particle on the flow behavior and the phenomenon
of particle migration through MRI techniques.

\appendix

\section*{Appendix A: Nonmechanical effects}\label{section_nonmechanical_effects}

In this appendix, we present how physicochemical interactions
between some particles and some pastes can be evidenced: first,
thanks to the measurement of the temporal evolution of the elastic
modulus, second, thanks to the experiments performed with various
bead sizes. This allows eliminating unambiguously some materials
from the analysis. This shows that all the points we raised in
Sec.~\ref{section_display} should be checked carefully when
embedding particles in a yield stress fluid if one wants to deal
with the general problem of any rigid particle in any yield stress
fluid.

\subsection*{Physicochemical effects in the emulsion and the
bentonite suspension}\label{section_chemical_effects_in_emulsion}

As mentioned in Sec.~\ref{section_elastic}, a condition necessary
to fulfill if there are no physicochemical interactions between
the beads and the yield stress fluid is that the dimensionless
elastic modulus $G'(\phi,t)/G'(0,t)$ does not depend on time,
whatever the volume fraction $\phi$ is. This condition was
fulfilled on the data presented on Fig.~\ref{fig6}.

\begin{figure}[htbp] \begin{center}
\includegraphics[width=9.5cm]{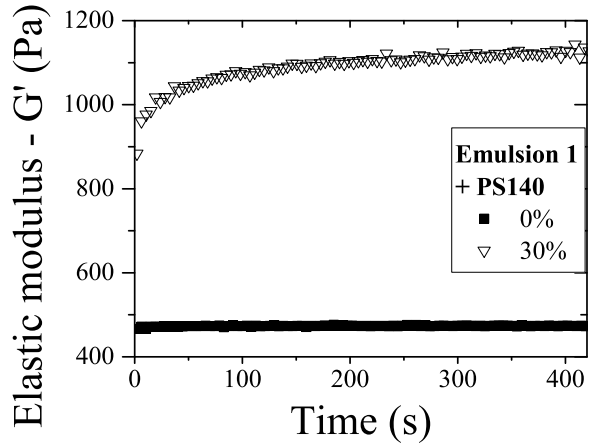}
\caption{Elastic modulus vs. time for a pure emulsion and for the
same emulsion filled with 30\% of 140$\mu$m PS
beads.}\label{elasticite_emulsion_modifiee_par_billes}
\end{center} \end{figure}

In a few cases, we found however that it would depend on time. In
the case of the bentonite suspension, one can still argue that our
procedure does not allow to get a reproducible initial state, and
that may cause $G'(\phi,t)/G'(0,t)$ to evolve in time as the
interstitial paste in the suspension is not the same as the pure
paste. In the case of the emulsion, the pure emulsion modulus does
not depend on time. Therefore, even if the manual preparation does
not allow to get a perfectly reproducible interstitial paste as
mentioned in Sec.\ref{section_display}, this paste properties
should not evolve in time when beads are embedded into it.
Fig.~\ref{elasticite_emulsion_modifiee_par_billes} shows the
elastic modulus evolution in time of an emulsion and of the same
emulsion filled with 140$\mu$m PS beads.

In Fig.~\ref{elasticite_emulsion_modifiee_par_billes} we observe
that the pure emulsion modulus is constant in time, whereas the
30\% suspension modulus evolves in time; there is no reason why it
should evolve, therefore we have to conclude that the emulsion
studied in Fig.~\ref{elasticite_emulsion_modifiee_par_billes} was
somehow modified by the inclusion of beads. Note that this
behavior was observed only in 2 emulsions, with 80 and 140$\mu$m
PS beads (even if carefully washed).  We did not correlate this
behavior with a particular formulation.

All the materials exhibiting this behavior were eliminated from
our analysis as they did not fulfill our aim of having purely
mechanical interactions.

More generally, looking at the dimensionless modulus as a function
of time thus seems a good test to study any suspension of
particles in a yield stress fluid if one seeks a general result
(i.e. not specific to the materials studied).

\subsection*{Size effects in Carbopol gels}
\label{carbopol_bead_size}

As mentioned in Sec.~\ref{section_elastic}, a condition necessary
to fulfill if there are no physicochemical interactions between
the beads and the yield stress fluid is that the dimensionless
elastic modulus $G'(\phi)/G'(0)$ does not depend on the bead size.
This condition was fulfilled on the data presented on
Fig.~\ref{fig8}.

\begin{figure}[htbp] \begin{center}
\includegraphics[width=13cm]{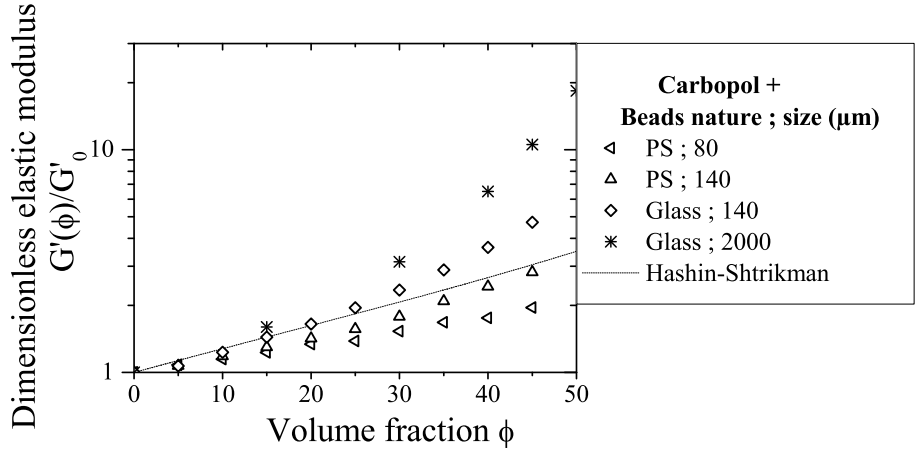}
\caption{Dimensionless elastic modulus $G'(\phi)/G'(0)$ vs. the
beads volume fraction $\phi$ for suspensions of 80$\mu$m,
140$\mu$m, and 2mm polystyrene and glass beads in a Carbopol gel.
The line is the Hashin-Shtrikman bound
$(2+3\phi)/(2-2\phi)$.}\label{carbopol_size_effect}
\end{center} \end{figure}

In Fig.~\ref{carbopol_size_effect}, we plot the dimensionless
elastic modulus $G'(\phi)/G'(0)$ vs. the beads volume fraction
$\phi$ for suspensions of 80$\mu$m, 140$\mu$m,  and 2mm
polystyrene and glass beads in a Carbopol gel.

We observe that in this material only, the evolution of the
dimensionless elastic modulus with the volume fraction depends on
the bead size and material. In the case of the 80$\mu$m and
140$\mu$m PS beads, the data fall below the Hashin-Shtrikman
bound: this means that the conditions are not fulfilled for
correct homogenization in these cases. On the other hand, as
expected, our data with the 2mm beads are consistent with all the
results obtained on the emulsion and the bentonite suspension,
whatever the bead size. This may be due to two problems: (i) the
Carbopol may be extremely sensitive to surface physicochemical
interactions; (ii) the microstructure size may be of the order of
several tens of $\mu$m (as mentioned in the introduction, the
Carbopol gel microstructure is actually poorly known and there is
only indirect evidence that the microstructure size may be of the
order of a few microns). In both cases, we expect the effect of
the size to disappear when the bead size is increased.

This problem, together with the structuration of Carbopol gels
under shear (see the end of Sec.~\ref{section_display}), point out
the fact that Carbopol gels are probably not the ideal model yield
stress fluids. It is important to note that this complexity exists
as Carbopol gels are often used as model yield stress fluids. We
think that an emulsion is more fitted to play this role.

\section*{Appendix B: Comparison with colloidal suspensions}

In order to see if the features we observe, and particularly
Eq.~\ref{eq_elasticity_vs_yield_stress}, are special features of
noncolloidal suspensions embedded in a colloidal paste, we have
qualitatively studied the evolution of the yield stress and the
elastic modulus of a colloidal paste with the volume fraction of
its colloidal particles. We prepared bentonite suspensions of
volume fraction $\psi$ ranging between 3 and 9\%, and measured the
evolution with time and concentration of their elastic modulus
$G'(\psi,t)$ and yield stress $\tau_c(\psi,t)$ with the procedures
defined in Sec.~\ref{section_display}.

\begin{figure}[htbp] \begin{center}
\includegraphics[width=8cm]{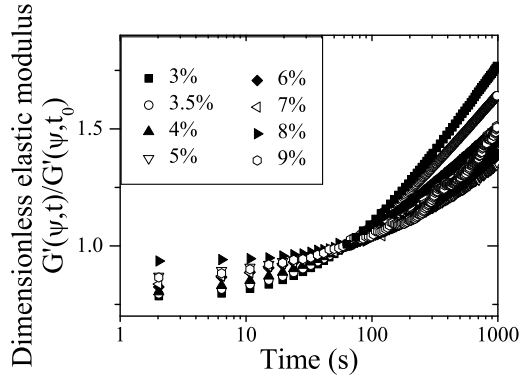}
\caption{a) Dimensionless elastic modulus
$G'(\psi,t)/G'(\psi,t_0=60s)$ vs. time for bentonite suspensions
at bentonite particles volume fraction ranging between 3 and
9\%.}\label{Gprime_adim_bento}
\end{center} \end{figure}

A first striking difference between both systems is that it is
difficult to study the evolution of the elastic modulus or the
yield stress of thixotropic materials with the volume fraction,
because the kinetics of their evolution depends on the volume
fraction of colloidal particles: when plotted vs. time,
$G'(\psi,t)/G'(\psi,t_0=60s)$ does not yield a unique
time-dependent function but depends on the concentration (see Fig.
\ref{Gprime_adim_bento}). While the structuration kinetics of a
given thixotropic material does not depend on the noncolloidal
particles added into it (see Sec.\ref{section_elastic}), it
depends a lot on the concentration of colloidal particles.

At a first glance, it thus seems difficult to analyze the
colloidal particles case. It has been shown however by
\citet{Ovarlez2007} that it is possible to distinguish between the
influence of $\psi$ on the elastic modulus evolution (which is
what we are searching) and on the structuration kinetics. It is
out of the scope of this paper to go further on this particular
point, and at this stage we consider that computing the
dimensionless elastic modulus and yield stress at a given time
provides nevertheless a rather good evaluation of the colloidal
particle concentration $\psi$ influence on the mechanical
properties of the colloidal paste.

\begin{figure}[htbp] \begin{center}
\includegraphics[width=13.2cm]{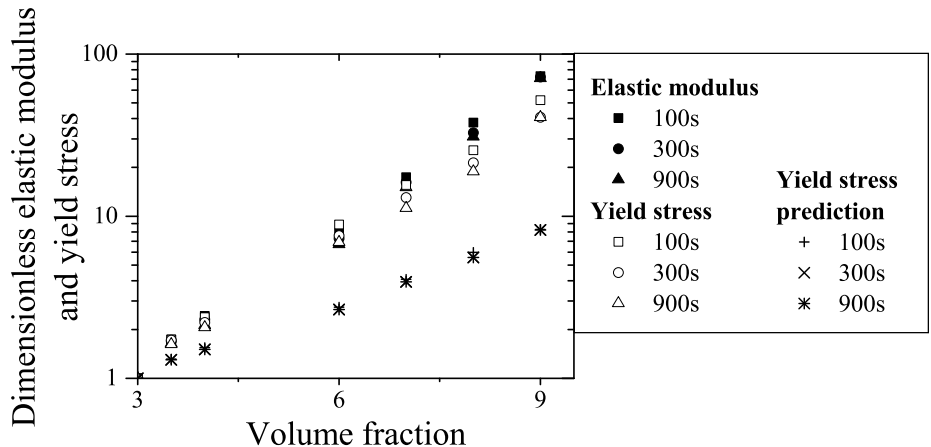}
\caption{Dimensionless elastic modulus
$G'(\psi,t)/G'(\psi_0=3\%,t)$ (filled symbols) and yield stress
$\tau_c(\psi,t)/\tau_c(\psi_0=3\%,t)$ (open symbols) computed at a
rest time $t=$100, 300 and 900s, vs. the colloidal particle volume
fraction $\psi$, for bentonite suspensions of bentonite particle
volume fraction ranging between 3 and 9\%. The crosses are the
dimensionless yield stress predictions from
Eq.~\ref{eq_elasticity_vs_yield_stress}.}\label{fig17}
\end{center} \end{figure}

In Fig.~\ref{fig17} we plot the dimensionless elastic modulus
$G'(\psi,t)/G'(\psi_0=3\%,t)$ and yield stress
$\tau_c(\psi,t)/\tau_c(\psi_0=3\%,t)$ computed at several rest
times $t$ vs. the colloidal particle volume fraction $\psi$ .

In contrast with what was observed when noncolloidal particles are
added into the paste (Sec.~\ref{section_elastic} and
Sec.~\ref{section_yield}), we observe that the yield stress and
elastic modulus evolution with $\psi$ are roughly identical. In
order to compare these results with the noncolloidal case, we also
plot in Fig.~\ref{fig17} the predictions for the yield stress
given by Eq.~\ref{eq_elasticity_vs_yield_stress} as a function of
the elastic modulus values: in the coordinates of
Fig.~\ref{fig17}, this would yield:
\bea\frac{\tau_{c,model}(\psi)}{\tau_{c,model}(\psi_0=3\%)}=\sqrt{\frac{(1-\psi)G'(\psi)}{(1-0.03)G'(\psi_0=3\%)}}\eea
We find that the model describing the influence of noncolloidal
particles would predict a much lower increase of the yield stress.

To conclude, the evolution of the behavior of suspensions of
noncolloidal particles in yield stress fluids with the
noncolloidal particle volume fraction $\phi$ is very different
from the evolution of the properties of a colloidal paste when
increasing the colloidal particle volume fraction $\psi$. The
colloidal paste structuration kinetics depends on $\psi$ (whereas
it does not depend on $\phi$), and its yield stress $\tau_c$
follows the same rapid evolution with $\psi$ than its elastic
modulus G' (whereas $\tau_c$ increases much less rapidly with
$\phi$ than G', see Eq. \ref{eq_elasticity_vs_yield_stress}).

\end{document}